\documentclass[a4paper]{article}
\pdfoutput=1

\usepackage[utf8]{inputenc}
\usepackage{graphicx}
\usepackage[type={CC},modifier={by},version={4.0},imagemodifier={-80x15},{hyperxmp=false}]{doclicense}

\usepackage{amsmath}
\usepackage{amssymb}
\usepackage{amsthm}
\usepackage{mathtools}
\newtheorem{definition}{Definition}
\newtheorem{example}{Example}

\newcommand{\inst}[1]{\unskip$^{#1}$}

\usepackage[hyphens]{url}
\usepackage[hidelinks]{hyperref}
\usepackage{cleveref}
\usepackage{todonotes}

\hypersetup{
    colorlinks=true,
    linkcolor=blue,
    citecolor=blue,
    urlcolor=violet,
    pdftitle={XTS mode revisited: high hopes for key scopes?},
}

\title{XTS mode revisited: high hopes for key scopes?}

\author{
Milan Brož\inst{1,2} \\ \normalsize\href{mailto:gmazyland@gmail.com}{gmazyland@gmail.com}
\and
Vladimír Sedláček\inst{2} \\ \normalsize\href{mailto:vlada.sedlacek@mail.muni.cz}{vlada.sedlacek@mail.muni.cz}
}

\date{
\vspace{0.5pt}
\normalsize\inst{1}Cryptsetup project \\
\normalsize\inst{2}Faculty of Informatics, Masaryk University, Czechia \\
\vspace{5pt}
\normalsize\today
}

\begin{document}
\maketitle

\begin{abstract}
This paper concisely summarizes the XTS block encryption mode for storage sector-based encryption applications and clarifies its limitations. In particular, we aim to provide a unified basis for constructive discussions about the newly introduced key scope change to the IEEE 1619 standard.  We also reflect on wide modes that could replace XTS in the future.
\end{abstract}

\begin{center}
\doclicenseText
\end{center}

\section{Introduction}

Since the introduction of XTS \cite{IEEE1619-2018} in 2007, many software and hardware sector-based encryption systems have widely adopted it, e.g., BitLocker \cite{bitlocker}, VeraCrypt \cite{veracrypt}, Cryptsetup \cite{cryptsetup}, and TCG Opal \cite{tcgopal} self-encrypting drives. Still, it seems to be surrounded by a shroud of misunderstandings, and some of its subtler points are still sometimes contested in the developer communities. It does not help that the existing NIST XTS-AES recommendation \cite{nistxts} only references the paid version of the IEEE standard \cite{IEEE1619-2018}, making it probably the only NIST document that does not include a publicly available definition of the described cryptographic primitive (as already mentioned in Rogaway's analysis \cite{rogaway2011evaluation} in 2011).

The situation is especially problematic in light of the newly introduced key scope change to the IEEE 1619-2025 standard \cite{IEEE1619-2025} (again behind a paywall), which would render the vast majority of implementations non-compliant. This would force vendors to request (not only) open-source projects to make significant changes to adapt to the new version. In contrast, the reasons and security implications of the change are not communicated at all.

The goals of this paper are threefold:
\begin{itemize}
    \item Unify the terminology in the XTS context in a way easily understandable to both theoreticians and practicioners (\Cref{sec:terminology} and \ref{sec:xts}).
    \item Describe XTS security limitations (namely key scopes, maximal sector size, and distinct keys) and clarify some contested points (\Cref{sec:limits}).
    \item Encourage a constructive public discussion to influence future requirements and recommendations (\Cref{sec:discussion}).
\end{itemize}

\section{Terminology and definitions} \label{sec:terminology}

\subsection{Units and sectors}\label{sec:hardware}
The storage industry is known to mix the power-of-10 units (SI) and the power-of-2 units for a storage size. To avoid any confusion, we strictly use powers of two with IEC prefixes: a byte is 8 bits, a kibibyte (KiB) is $2^{10}$ bytes, a mebibyte (MiB) is $2^{20}$ bytes (1024~KiB), and a tebibyte (TiB) is $2^{40}$ bytes (1024~MiB).

A block device is an abstraction of either a physical device (like a hard disk or a flash-based device) or a virtualized device that can be accessed (read, written to, or erased) only in units called sectors. A typical sector size is 4096 bytes (4~KiB) or 512 bytes (for older devices).
In this text, we will only consider encryption of block devices (even though XTS can be used in other scenarios).
Furthermore, we assume that the sector is always read or written to atomically, and individual sectors are accessed independently.\footnote{Modern storage devices are complex systems that can internally use different storage blocks for optimal performance, and can have specific requirements for optimal operations.}

The typical blocksize for modern blockciphers is much smaller than the sector size (16 bytes for AES).
Thus, to encrypt the whole sector, we need to use an encryption mode (such as XTS) that builds upon the original cipher. All modern real-world storage devices always use sector sizes that are multiples of 16 bytes, eliminating the need for padding during both encryption and decryption.\footnote{The only legacy exception is the 520-byte sector (a 512-byte data sector with an additional 8-byte integrity field) on enterprise devices.}

\subsection{Notation}
In this text, $E$ will denote a blockcipher (e.g., AES-128) of blocksize~$n$ bits, so that for each key $K$, $E_K$ is a permutation on $\{0,1\}^n$ -- the set of binary $n$-bit strings. For $s,s'\in \{0,1\}^n$, we will write $s \Vert s'$ and $s \oplus s'$, for the concatenation and XOR operations, respectively.
A binary string with bits $a_{n-1},\dots, a_1, a_0$ corresponds to the element $[a_{n-1}x^{n-1} + \dots + a_1x + a_0]$ of the finite field $\mathbb{F}_{2^n}$.
Thus, we can multiply binary strings by elements of $\mathbb{F}_{2^n}$ using shift and XOR operations (though it depends on the polynomial chosen to represent $\mathbb{F}_{2^n}$).

$S$ will denote the total number of sectors on the device and $J$ the number of $n$-bit blocks within a sector. Correspondingly, $N$ will always denote a sector number and $j$ the number of the $n$-bit block within the sector.

We will denote the unencrypted data (plaintext) at block $j$ within sector $N$ by $P_{N,j}$ and naturally extend this by concatenation:
$$P_N := P_{N,0} \Vert P_{N,1} \Vert \dots \Vert P_{N,J-1}$$
will denote the unencrypted data at sector number $N$, while
$$P := P_0 \Vert P_1 \Vert \dots \Vert P_{S-1}$$
will denote the unencrypted data of the whole device. Analogically, we denote the encrypted data by $C_{N,j}:=E_K(P_{N,j}), C_n, C$, respectively.

\section{XTS encryption mode} \label{sec:xts}
The XTS mode -- or rather XTS-AES -- was first defined in Annex D.4.3 in \cite{IEEE1619-2018} as an ``instantiation'' of Rogaway's XEX \cite{rogaway2004efficient} with the underlying blockcipher being AES-128 or AES-256. Though not quite, as there are several differences:
\begin{itemize}
    \item While XEX uses a single symmetric key to encrypt both the data and the sector number, XTS uses two different symmetric keys: $K$ to encrypt the data, and $K_T$ to encrypt the sector number.
    \item While XEX starts at $j=1$, XTS starts at $j=0$.
    \item XTS allows for ciphertext stealing to deal with encrypting data whose size is not a multiple of $n$. As explained in \Cref{sec:hardware}, we focus only on full sector encryption, so we can ignore this.\footnote{XTS has been used in the Linux kernel \href{https://git.kernel.org/pub/scm/linux/kernel/git/torvalds/linux.git/commit/crypto/xts.c?id=f19f5111c94053ba4931892f5c01c806de33942e}{since 2007}.
    In contrast, ciphertext stealing was only added \href{https://git.kernel.org/pub/scm/linux/kernel/git/torvalds/linux.git/commit/crypto/xts.c?id=8083b1bf8163e7ae7d8c90f221106d96450b8aa8}{in 2019},
    and probably there is still no active in-kernel user.
    }
\end{itemize}

\subsection{XTS definition}\label{sec:xts-definition}
We propose a definition that seeks a compromise between Rogaway's original XEX formality and the practical use case at hand.
\begin{definition}[XTS encryption mode]
    Let $K$, $K_T$ be symmetric keys for a blockcipher $E$ of blocksize\footnote{This is not essential, but multiplication by $\alpha$ differs for different block sizes.} 128 bits. Then the XTS mode of $E$ encrypts the storage data in the following way:
    $$C_{N,j} :=
    E_K(P_{N,j} \oplus T_{N,j}) \oplus T_{N,j},$$
    $$C_N := C_{N,0} \Vert C_{N,1} \Vert \dots \Vert C_{N,J-1},$$
    $$C := C_0 \Vert C_1 \Vert \dots \Vert C_{S-1},$$
    where $$T_{N,j} := E_{K_T}(N) \cdot \alpha ^ j$$
    and $\alpha$ is the element $[x]$ in $\mathbb{F}_{2^{128}}:= \mathbb{F}_{2}[x]/(x^{128} + x^7 + x^2 + x + 1)$.
\end{definition}

Conceptually, the multiplication by $\alpha$ corresponds to a left shift with overflow. More explicitly, for a binary string $s = a_{127}a_{126}\dots a_1a_0$ (this is how we denote concatenation of bits), we have
\[ s\cdot \alpha =
    \begin{cases*}
        a_{126}a_{125}\dots a_0 0 & if $a_{127} = 0$,  \\
        a_{126}a_{125}\dots a_0 0 \oplus \underbrace{0\dots 0}_{120}10000111 & if $a_{127} = 1$,
    \end{cases*} \]%
as $x^{128}$ and $x^7 + x^2 + x + 1$ represent the same element in
$\mathbb{F}_{2^{128}}$.
\newpage

\begin{figure}[ht]
\begin{center}
    \includegraphics[width=0.88\linewidth]{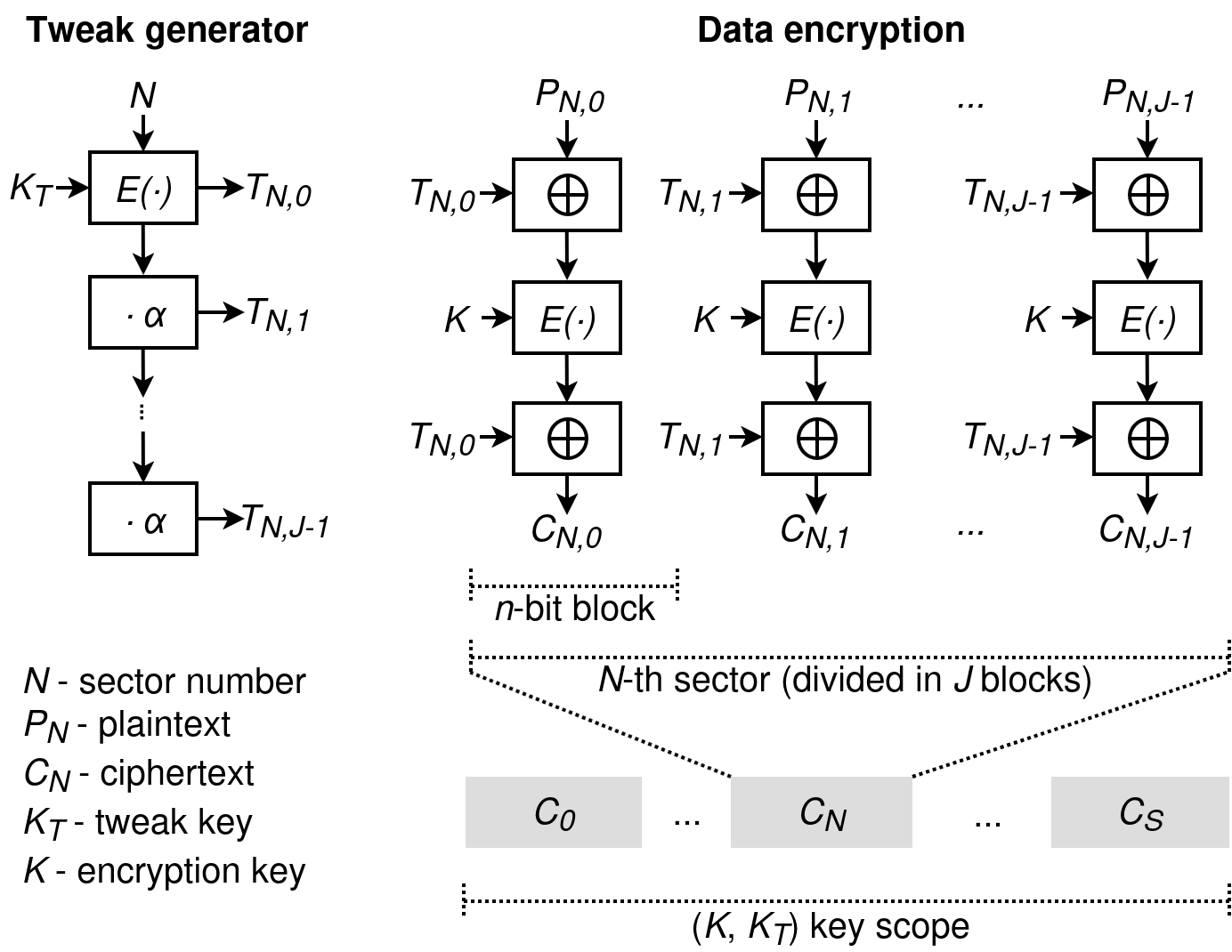}
    \caption{The principle of XTS sector encryption (without ciphertext stealing).}
    \label{fig:xts}
\end{center}
\end{figure}

\begin{example}[AES-XTS-128 encryption of one sector with two blocks]\hfill\break
\indent For
\begin{align*}
N       & = 0x00000000000000000000000000000001, \\
K       & = 0x11111111111111111111111111111111, \\
K_T     & = 0x22222222222222222222222222222222, \\
P_{N,0} & = 0x44444444444444444444444444444444, \\
P_{N,1} & = 0x88888888888888888888888888888888,
\end{align*}

the tweak values are generated in the following way:
\begin{align*}
T_{N,0} & = E_{K_T}(N)             = 0x6752ca5febca0f3fc8dc9dfc2a916295, \\
T_{N,1} & = T_{N,0} \cdot \alpha^1 = 0x49a494bfd6951f7e90b93bf95522c52a.
\end{align*}

The sector blocks are then encrypted as follows:
\begin{align*}
C_{N,0} & = E_K(P_{N,0} \oplus T_{N,0}) \oplus T_{N,0} \\
        & = E_K(0x6752ca5febca0f3fc8dc9dfc2a916295) \oplus T_{N,0} \\
        & = 0x13f084e65a7ca361f74957c9b11c7710 \oplus T_{N,0} \\
        & = 0x74a24eb9b1b6ac5e3f95ca359b8d1585 \\
\\
C_{N,1} & = E_K(P_{N,1} \oplus T_{N,1}) \oplus T_{N,1} \\
        & = E_K(0xc12c1c375e1d97f61831b371ddaa4da2) \oplus T_{N,1} \\
        & = 0x2cada9d22ad34bf19a226c2c824f0364 \oplus T_{N,1} \\
        & = 0x65093d6dfc46548f0a9b57d5d76dc64e
\end{align*}

\end{example}

\clearpage
\subsection{Threat model}
The simplest threat model for length-preserving sector encryption usually considers a ``stolen device'', allowing the attacker to access only one ciphertext snapshot at time.
A typical example of such a threat model is the TCG Opal definition \cite{tcgopal}:
\emph{``protect the confidentiality of stored user data against unauthorized access once it leaves the owner's control (following a power cycle and subsequent deauthentication).''}
This model expects the user to recognize when the protected device ``left the owner's control'', which is not always feasible.

The real-world threat model also considers situations where the attacker can manipulate the ciphertext after repeatable access to the encrypted device, even performing some form of ``traffic analysis''. This is important, as encrypted images are often stored in cloud environments.
Section D.4.3 in \cite{IEEE1619-2025} suggests (in addition to introducing a key scope): \emph{``The decision on the maximum amount of data to be encrypted with a single key should take into account the above calculations together with the practical implication of the described attack (e.g., ability of the adversary to modify plaintext of a specific block, where the position of this block may not be under adversary’s control).''}.

Relatedly, we should consider adversaries capable of the chosen-ciphertext attack (CCA), i.e., able to query both the encryption and decryption oracle (with $K, K_T$ still being fixed and secret) on inputs of their choice.

Despite having received a lot of criticism \cite{rogaway2011evaluation,ptacek2014xts}, the XTS mode has not been replaced by any other mode more than a decade later. Unfortunately, its security goals and threat models are still not well-defined.

\section{Security limits of XTS}\label{sec:limits}

\subsection{Key scopes}\label{sec:key-scopes}
The most drastic change in the IEEE 1619-2025 standard \cite{IEEE1619-2025} is the following requirement: ``\emph{Maximum Number of 128-bit blocks in a key scope = $2^{36}$ to $2^{44}$}''. For a fixed XTS key\footnote{By an XTS key, we really mean a concatenated pair of symmetric keys; this is standard terminology.} $(K, K_T)$, this translates to $S\cdot J \le 2^{36}$ and $S\cdot J \le 2^{44}$, respectively.

The reason for this comes from Annex D.4.2 in \cite{IEEE1619-2025}, which considers a CCA adversary with access to triples $(P_{N,j}, C_{N,j}, T_{N,j})$ and $(P_{N',j'}, C_{N',j'}, T_{N',j'})$ for some $N, N', j, j'$, such that
\begin{equation}\label{collision}
    P_{N,j} \oplus T_{N,j} = P_{N',j'} \oplus T_{N',j'}.
\end{equation}

If such an adversary can overwrite $P_{N,j}$, they can use the resulting ciphertext to change $C_{N',j'}$ to a value that decrypts to any plaintext of their choice.

There is a non-negligible probability that \eqref{collision} occurs when the number of plaintexts ($S\cdot J$) encrypted under the same XTS key $(K, K_T)$ approaches the birthday bound $2^{n/2}$.

\clearpage
Consequently, the scope of each key should be limited. The probability of \eqref{collision} occurring for AES-\{128,256\} can be estimated by $\frac{(S\cdot J)^2}{2^{128}}$. Thus, encrypting 1 TiB of data (i.e., $S\cdot J=2^{36}$) would lead to \eqref{collision} occurring with probability roughly $2^{36\cdot 2-128}=2^{-56}$, whereas 256 TiB of data would correspond to roughly $2^{44\cdot 2-128}=2^{-40}$. This agrees with Annex D.4.3 in \cite{IEEE1619-2018} up to a small constant.\footnote{
Annex D.4.3 in \cite{IEEE1619-2018} mistakenly uses a general security guarantee $9.5 \frac{q^2}{2^n}$ from \cite{rogaway2004efficient} (valid for XEX, not XTS), but in fact, the factor $9.5$ can be dropped if AES is secure \cite{liskov2008comments,ball2012xts}.
}
However, it is unclear what amount of risk is acceptable in different contexts,\footnote{We discussed this issue both in an IEEE SISWG meeting on April 25, 2025 and associated private communication, and got a confirmation that the analysis \ref{sec:key-scopes} is the one motivating the key scopes. The goal was to satisfy
security targets set by NIST (i.e., $2^{-53}$ and $2^{-37}$, respectively, should be good enough), but NIST never explained the choice of the specific constants, just like with the maximal sector size in \Cref{sec:maxsectorsize}.} even though the change would invalidate compliance of many current  encrypted storage devices.

\subsection{Maximal sector size}\label{sec:maxsectorsize}
Section 5.1 in \cite{IEEE1619-2018} states: ``\emph{The number of 128-bit blocks within the data unit shall not exceed $2^{20}$.}'' As their data unit translates to our sector, this limits its size to 16 MiB, i.e., $J \le 2^{20}$. This condition has already been present in \cite{IEEE1619-2018} as a change from the initial version (2007), where it was only a suggestion (``\emph{should not exceed}''). NIST \cite{nistxts} also strictly mandates this limit, but the rationale has never been explained. As XEX has been proven to be secure against chosen-ciphertext attacks even without this condition \cite{rogaway2004efficient,minematsu2006improved}, its purpose remains unclear.
In practice, this is almost never an issue, as the typical sector size is at most 4~KiB, i.e., $J \le 2^8$ (\Cref{sec:hardware}).

\subsection{Distinct \texorpdfstring{$K, K_T$}{K, KT}}
Annex C.I in \cite{fips-140-3} warns about a chosen ciphertext attack against XTS-AES, referring to Section 6 in Rogaway's original paper \cite{rogaway2004efficient}: ``\emph{by
obtaining the decryption of only one chosen ciphertext block in a given data sector, an adversary who does not
know the key may be able to manipulate the ciphertext in that sector so that one or more plaintext blocks
change to any desired value}''. The attack crucially relies on two simultaneous conditions: $K=K_T$,
and starting the indexation at $j=0$ (where the tweak is unchanged, as it is multiplied by $\alpha^0$).
There are two obvious ways to prevent the attack:
\begin{itemize}
    \item Starting from $j=1$ (as XEX does); then the original security proof applies (Section 6 in \cite{rogaway2004efficient}, Section 4.2 in \cite{minematsu2006improved}).
    \item Requiring that $K \ne K_T$. FIPS 140-3 \cite{fips-140-3} made this option mandatory for compliance, even though as Liskov and Minematsu argue in \cite{liskov2008comments}, this comes from a misapplication of a security design practice (though does not lower the security in any way).
\end{itemize}

\section{Discussion} \label{sec:discussion}
The XTS encryption mode, as defined in IEEE 1619 \cite{IEEE1619-2018}, has been used for many years in many different systems. Thus, any compliance-breaking changes -- such as the mandatory use of key scopes in the IEEE 1619 standard \cite{IEEE1619-2025} -- could have a dramatic impact on the ecosystem. This would be amplified even more by the concept of FIPS certifications (which should impact the document \cite{fips-140-3}).

We are not necessarily arguing against the change; the issue described in \Cref{sec:key-scopes} could definitely be a substantial one in some scenarios. However, to justify the increased complexity and potential for new implementation mistakes, it would be helpful to first have good answers to the following questions:
\begin{itemize}
    \item What are the contexts with a weaker threat model? For example, in the ``stolen'' device model, where the attacker cannot actively modify the ciphertexts, adding key scopes would not improve security.
    \item What are the contexts with a stronger threat model? For example, if the attacker can store or extract snapshots of encrypted blocks outside the device, adding key scopes would not necessarily protect against the attack.
    \item Is there another way to strengthen XTS without the key scopes? One option would be using a blockcipher with larger $n$. As $n=256$ would already be acceptable, and NIST already plans to propose a standard for ``wide AES'' (\cite{nistwideaes}), this option should be included in future considerations. The problem is that the XTS standard \cite{IEEE1619-2018} has hardcoded 128-bit blocks everywhere and would need to be updated.
    \item How should the XTS keys used for key scopes be generated? It is unclear if all the keys used in key scopes must be generated using an approved random number generator, or if they can be derived from a master key with an approved key derivation algorithm. The second option would simplify key scope implementation.
    \item How to determine which XTS key to use for which sector? And how to handle device resizing? Standardizing (several variants of) answers to these might help with complexity reduction.
\end{itemize}

Currently, XTS-AES uses a single XTS key for the whole device (for all addressable sectors). Introducing key scopes can lead to incompatible variants of XTS-AES implementations until we specify which XTS key corresponds to a given sector. We propose two partial answers to this problem:
\begin{itemize}
    \item Implement key scopes in a linear fashion: use each XTS key to encrypt the next (at most) $2^{44}$ plaintexts, then switch to the next key. A downside is that keys would not be utilized uniformly: if data is stored only on a small part of the device, the keys beyond the first one are never used. Also, for device resizing, keys would need to be added or removed dynamically.

    \item Implement key scopes with rotating XTS keys: use $(N \text{ mod } m)$-th XTS key to encrypt the $N$-the sector, where $m$ is the total number of XTS keys.
    This solution must define the maximal device size in advance (to not exceed the required key scope), but easily allows a dynamic resize (up to the maximal size). It also utilizes all keys more uniformly.
\end{itemize}

\subsection*{Beyond XTS mode}
We would like to stress that XTS comes with serious downsides, just like other classical modes such as ECB, CBC, CFB, OFB, and CTR. Namely, they fail to provide full diffusion (i.e., each bit of the ciphertext depending on each bit of the plaintext) if the data to be encrypted exceeds a few blocks \cite{dobraunig2025efficient}.

For a long-term use of a length-preserving mode (i.e., disqualifying authenticated encryption), it seems much more fruitful to switch to a wide encryption mode suited for disk encryption (as suggested in \cite{rogaway2011evaluation}). The feasible candidates might include:
\begin{itemize}
    \item EME2 is based on EME, designed in 2003 and later standardized by IEEE 1619.2 \cite{eme-ieee}. The mode was never widely adopted, partially because of a questionable status of the related patent (now abandoned\footnote{\url{https://lore.kernel.org/dm-crypt/kf5lug$8p0$1@ger.gmane.org/}}).
    \item Adiantum \cite{crowley2018adiantum} was developed by Google for low-end systems that do not provide AES hardware acceleration. It is designed to be fast and has low power requirements (suitable for mobile devices).
    \item HCTR2 \cite{crowley2021hctr2} (also by Google) builds upon the CTR mode and seems like a performant and conservative option.
    \item \emph{bbb-ddd-AES} \cite{dobraunig2025efficient} is a new mode built over the docked-double-decker construction \cite{gunsing2019deck}. If there is an upper bound on the number of times each tweak can be reused (sometimes satisfied by SSD hardware due to limited lifetime and wear-leveling), it can provide beyond-birthday bound security. The double-decker construction is a flexible framework that allows building wide encryption modes with better characteristics than the older modes mentioned above. As the proposed modes are very recent, there is no independent security analysis, and -- unlike for the above candidates -- no public reference implementation.
\end{itemize}

The pragmatic solution is perhaps to keep XTS limits in a sustainable state, as many legacy systems will be using it in the next years. For long-term improvement of disk encryption, we believe it would be worthwile to focus on a newer construction, like the mentioned double-decker framework.

We would be grateful for any feedback or suggestions that could contribute to the improvement of this work. The source code of this document is available on GitHub\footnote{\url{https://github.com/mbroz/xts-paper}}.

\section*{Acknowledgements}
The authors were supported by the European Union under Grant Agreement No.~101087529 (CHESS -- Cyber-security Excellence Hub in Estonia and South Moravia).

\clearpage

\bibliographystyle{plain}
\bibliography{xts-paper}

\end{document}